\documentclass[camera,letterpaper,nomarginnotes,nonarrowgutter]{jpaper}

\usepackage{xcolor}
\definecolor{freakishgreen}{HTML}{0A982B}
\definecolor{urlblue}{HTML}{319dd6}
\usepackage[colorlinks=true, linkcolor=blue, citecolor=violet, urlcolor=urlblue]{hyperref}
\usepackage{dblfloatfix}
\usepackage{multirow}
\usepackage{booktabs}
\usepackage{array}
\usepackage{algorithm}
\usepackage{eucal}
\usepackage{fancyhdr}
\usepackage{subfigure}
\usepackage{amsmath,amssymb,amsfonts}
\usepackage{graphicx}
\usepackage[dvipsnames]{xcolor}
\usepackage[auth-lg,affil-it]{authblk}

\usepackage{blindtext,graphicx}
\usepackage[absolute]{textpos}
\usepackage{enumitem}
\usepackage[noend]{algpseudocode}
\newcolumntype{L}[1]{>{\raggedright\let\newline\\\arraybackslash\hspace{0pt}}m{#1}}
\newcolumntype{C}[1]{>{\centering\let\newline\\\arraybackslash\hspace{0pt}}m{#1}}
\newcolumntype{R}[1]{>{\raggedleft\let\newline\\\arraybackslash\hspace{0pt}}m{#1}}
\usepackage{tikz}

\makeatletter
\def\thickhline{%
  \noalign{\ifnum0=`}\fi\hrule \@height \thickarrayrulewidth \futurelet
   \reserved@a\@xthickhline}
\def\@xthickhline{\ifx\reserved@a\thickhline
               \vskip\doublerulesep
               \vskip-\thickarrayrulewidth
             \fi
      \ifnum0=`{\fi}}
\makeatother
\newlength{\thickarrayrulewidth}
\usepackage{pifont}

\usepackage{cleveref}
\crefname{section}{§\hspace{-2pt}}{§§}
\Crefname{section}{§}{§§}
\usepackage{datetime}

\usepackage[compress,sort]{cite}

\newif\ifsubmission
\submissiontrue

\definecolor{brickred}{rgb}{0.8, 0.25, 0.33}
\definecolor{darkblue}{rgb}{0, 0, 0.5}  

\definecolor{darkyellow}{rgb}{0.6, 0.6, 0}  

\definecolor{somecolor}{rgb}{0.21, 0.51, 0.22}
\definecolor{purple}{rgb}{0.4, 0.0, 0.6}
\definecolor{brown}{rgb}{0.6, 0.2, 0.2}

\newcommand{\juan}[1]{\textcolor{black}{#1}}
\newcommand{\juang}[1]{\textcolor{black}{#1}}
\newcommand{\hg}[1]{\textcolor{black}{#1}}
\newcommand{\mk}[1]{\textcolor{black}{#1}}
\newcommand{\hgupta}[1]{\textcolor{black}{#1}}
\newcommand\ignore[1]{ }
\newcommand{\guptah}[1]{\textcolor{black}{#1}}
\newcommand{\guptahh}[1]{\textcolor{black}{#1}}
\newcommand{\guptahhh}[1]{\textcolor{black}{#1}}
\newcommand{\mkk}[1]{\textcolor{black}{#1}}
\newcommand{\hgg}[1]{\textcolor{black}{#1}}
\newcommand{\hggg}[1]{\textcolor{black}{#1}}

\newcommand{\tklbl}[0]{Key Takeaway}
\newcounter{take}
\newcommand\takeaway[1]{%
   \stepcounter{take}
   \noindent
   \colorbox{gray!20}{\textbf{\tklbl{} \thetake.}} \emph{#1}}
    
    \fancypagestyle{firstpage}
    {
        \fancyfoot[C]{\thepage}
    }

\newif\ifarxiv
\arxivfalse

\usepackage{flushend} 

\usepackage{xspace}

\begin{document}
\urlstyle{tt}
\bstctlcite{IEEEexample:BSTcontrol}
%
\title{Evaluating Homomorphic Operations \\ on a Real-World Processing-In-Memory System}


\author{
Harshita Gupta$^*$\hspace{1em}
Mayank Kabra$^*$\hspace{1em}
Juan Gómez-Luna\hspace{1em} 
Konstantinos Kanellopoulos \hspace{1em} 
Onur Mutlu \vspace{-0.5em} \\
\normalsize{ETH Zürich}
}

\maketitle


\thispagestyle{firstpage}

\begin{abstract} \label{sec:abstract}
\hgupta{Computing on encrypted data is a promising approach to reduce data security and privacy risks, with \mkk{homomorphic} encryption serving as a facilitator in achieving this goal.}
\juan{In this work, }\mk{we accelerate} \guptahh{homomorphic operations} \mk{using the Processing-in-Memory (PIM)} \mk{paradigm}  \mk{to mitigate the large memory capacity and frequent data movement requirements.} 
Using a real-world PIM system, 
we accelerate \juan{the} Brakerski-Fan-Vercauteren (BFV) scheme 
\juan{for} \hgupta{homomorphic} addition \juan{and} multiplication. 
\juan{We evaluate the PIM implementations of these \guptahh{homomorphic} operations with statistical workloads (arithmetic mean, variance, linear regression) and compare to CPU and GPU implementations.} 
\mk{Our results demonstrate} 
$50-100\times$ speedup with a real PIM system (UPMEM) over the CPU and $2-15\times$ over the GPU in vector addition.
\hgupta{For vector multiplication, the real PIM system outperforms \mk{the} CPU by $40-50\times$. However, it}
lags $10-15\times$ behind the GPU due to \juan{the lack of native} \mk{sufficiently wide multiplication support} \hgupta{in the evaluated first-generation real PIM system}. 
For mean, variance, and \juan{linear} regression, \mk{the real PIM system performance} improvements vary \mk{between} $30\times$ and $300\times$ over the CPU and \mk{between} $10\times$ \mk{and} $30\times$ over the GPU, uncovering \mk{real} PIM \mk{system} trade-offs \hgupta{in terms of scalability of \guptahh{homomorphic} operations for varying amounts of data.} \guptahh{We plan to make our implementation open-source in \guptahhh{the} future.}
\end{abstract}


%

\def\thefootnote{*}\footnotetext{Equal contribution.}\def\thefootnote{\arabic{footnote}}

\section{Introduction} \label{sec:introduction}

\hgupta{Traditional security measures \guptah{that operate on plain (unencrypted) data} often expose \guptah{the actual} data during processing, creating \guptah{security and privacy} vulnerabilities.
Homomorphic Encryption (HE) \cite{ogburn2013homomorphic, tourky2016homomorphic, gentry2011implementing, al2020towards, gentry2009fully, van2010fully, cryptoeprint:2014/356, boneh2013private} addresses this by enabling calculations on encrypted data without revealing sensitive information.}




\juan{A user can (1) encrypt data, and (2) send it to \mk{the server}. Then, (3) computing resources in the \mk{server} operate on the data without decrypting it, \guptahhh{using HE}, and (4) the encrypted results} \mk{are returned} to the user, preserving data privacy~\cite{moore2014practical, chaudhary2019Analysis, gupta2021accelerating, gentry2009fully}. 
However, HE 
\juan{is very costly due to the use of} large ciphertexts and \mkk{computation} intensive operations \cite{samardzic2021f1, samardzic2022craterlake, alaya2020homomorphic, el2015challenges}. For example, performing \guptahh{homomorphic} multiplication on two fully homomorphic (FHE) encrypted integers may require tens of millions of operations \cite{gupta2022memfhe, cao2015optimised, su2022highly}. 
The complexity is further compounded by intricate mathematical operations, as each of these operations is executed on data that can be up to $1000\times$ larger \mk{in size} than the original \guptah{plain} data \cite{gupta2022memfhe, tourky2016homomorphic, doroz2016homomorphic}.



\juan{Recent research proposes the implementation of \hgg{homomorphic} operations} 
on CPUs \cite{cao2015optimised, mert2019design, meftah2022towards, morshed2020cpu}, GPUs \cite{al2020towards, dai2016cuhe, al2020multi, morshed2020cpu, dai2014accelerating}, FPGAs \cite{wang2013fpga, agrawal2023fab, cousins2016designing, cousins2014fpga, syafalni2022efficient, jayet2015polynomial}, and ASICs \cite{feldmann2021f1, samardzic2022craterlake, yang2023poseidon, ozturk2016custom, kim2022bts, nabeel2023cofhee, cao2014high}, 
\juan{but these implementations
\guptah{do not}
fundamentally solve the \emph{data movement bottleneck} \mk{associated with} 
\guptahh{homomorphic} operations}.



\emph{Processing-in-Memory (PIM)}, 
\juan{i.e., equipping memory with compute capabilities}~\cite{gomez2021benchmarking, gomez2022benchmarking, kim2021aquabolt, kim2022aquabolt, lee2022improving, ke2021near, upmemsdk, gomez2022machine, item2023transpimlibispass, kim2014hbm, ghose.bookchapter19, oliveira2021damov, devaux2019true, ghose2019processing, mutlu2022modern, seshadri2017ambit, stone1970logic, Kautz1969, farmahini2014drama, singh2018review, kwon2019tensordimm, yitbarek2016exploring, gupta2022memfhe, hajinazar2021simdram, lim2023design}, can effectively alleviate \mk{the} data movement needs. 
Recent PIM-based HE solutions \cite{gupta2022memfhe, gupta2021accelerating, li2019leveraging, reis2020computing} leverage \juan{high parallelism and memory bandwidth} \guptah{inside the memory chips} 
for acceleration. 
\juan{
\guptah{However, there}
is no evaluation of \guptahh{homomorphic} operations on real PIM systems, which have recently
\hgg{been introduced}~\cite{gomez2021benchmarking, gomez2022benchmarking, kim2021aquabolt, kim2022aquabolt, lee2022improving, ke2021near, upmemsdk, gomez2022machine, item2023transpimlibispass, kim2014hbm, devaux2019true}.}

\juan{To our knowledge, this study is \mk{the} first to implement and evaluate \guptahh{homomorphic} operations on a real PIM system.} \guptah{Using a real PIM system (UPMEM)\cite{gomez2021benchmarking, gomez2022experimental, upmemsdk}\guptahhh{,}
we accelerate the Brakerski-Fan-Vercauteren (BFV) scheme~\cite{halevi2019improved, wibawa2022bfv} 
for homomorphic addition and multiplication.}
\juan{Our evaluation shows that \mk{the real PIM system} accelerates \guptah{the} homomorphic addition \mk{operation} by $50-100\times$ over \hgupta{a state-of-the-art} CPU and by $2-15\times$ over \hgupta{a state-of-the-art} GPU. 
For \guptah{the} homomorphic multiplication \mk{operation}, \mk{the real PIM system} provides a speedup of $30-50\times$ over \hgupta{the} CPU, but lags $10-15\times$ behind \hgupta{the} GPU due to the lack of native \hgupta{sufficiently wide} multiplication support on \hgupta{the evaluated first-generation} UPMEM PIM \mk{system}. 
\guptahh{We also evaluate our implementation of three statistical workloads (mean, variance, linear regression) using \guptahhh{homomorphic addition and homomorphic multiplication.} 
\guptahhh{In our evaluation, }the real PIM system achieves up to $300\times$ speedup over the CPU for all workloads and up to $30\times$ over the GPU for arithmetic mean. However, it lags by up to $50\times$ compared to the GPU for variance and linear regression, \guptahhh{due to the low performance of multiplication on the \mkk{first real-world} PIM system.}
}}

Our work makes the following contributions:
\begin{itemize}
\item \juan{We develop the first implementation of \guptahhh{homomorphic} addition and multiplication on a real PIM system.}

\item \juan{We evaluate the performance of \guptahhh{homomorphic} addition and multiplication on a real PIM system for different bit-key security levels \hg{(27-109 bits)}. We use three real-world statistical workloads (arithmetic mean, variance, linear regression) for evaluation.} 


\item Our findings demonstrate the \hgupta{capabilities}
\hgupta{and tradeoffs} of 
\juan{real PIM systems} for efficient 
cryptographic operations, providing a foundation for future developments in this 
\juan{direction}.



\end{itemize}

\section{Background and Motivation} \label{sec:motivation}

Homomorphic encryption (HE)\cite{ogburn2013homomorphic, tourky2016homomorphic, gentry2011implementing, al2020towards, gentry2009fully, van2010fully, cryptoeprint:2014/356, boneh2013private} 
enables processing \juan{(e.g., addition, multiplication, rotation)}  \guptah{on} encrypted data while preserving privacy. 
\juan{We focus} on 
\juan{the} BFV (Brakerski-Fan-Vercauteren) scheme \guptah{for HE}~\cite{halevi2019improved, wibawa2022bfv}, 
\juan{but the implementation techniques that we propose} 
are \guptah{also} applicable to other HE schemes 
\juan{(e.g., BGV \cite{mono2022finding} and CKKS \cite{cheon2020remark})}. 
HE types include Fully Homomorphic Encryption (FHE), Partially Homomorphic Encryption (PHE), and Somewhat Homomorphic Encryption (SHE)~\cite{acar2018survey}. 
FHE enables unrestricted operations, PHE permits one type of operation, and SHE supports both addition and multiplication with constraints on multiplicative depth. 
\juan{FHE, SHE, and PHE offer different trade-offs between security and efficiency~\cite{alexandru2020cloud, damgaard2012multiparty, boneh2013private, ogburn2013homomorphic, yasuda2014practical, gentry2009fully, cryptoeprint:2014/356}.}  
In this paper, we focus on SHE as it provides a balance between security and efficiency, allowing some computations \juan{
\guptah{(e.g.,} addition, multiplication)} 
on encrypted data while still maintaining a high level of security.

HE 
\juan{poses two main \textbf{challenges}} that limit its use in real-world applications\guptah{.}

\noindent\textbf{1) 
\juan{Large memory footprint}:} HE schemes 
\juan{require} very long vectors with wide elements to encode information~\cite{samardzic2022craterlake}. 
\juan{Prior work~\cite{feldmann2021f1} shows} that multiplying 2MB 
ciphertexts 
requires 32MB 
of auxiliary data, and 
25MB 
ciphertexts would require over 1.4GB 
of auxiliary data. This amount of auxiliary data is too large to fit 
\guptah{on} \mkk{a processor-centric chip} \juan{which limits \guptah{the} scalability \guptah{and performance} of HE}. 


\noindent\textbf{\juan{2) Frequent data} movement:} 
\juan{The large amount of data that \hgg{homomorphic} algorithms need to operate on is moved back-and-forth between off-chip memory/storage units and compute units. Prior work~\cite{de2021does} shows that homomorphic operations exhibit low arithmetic intensity (<1 operations/byte). 
As a result, \guptah{in processor-centric systems, such as CPUs and GPUs, it is challenging to efficiently offset the performance and energy expenses incurred when transferring large amounts of data.}}



\guptahh{Several recent works~\cite{al2020towards, dai2016cuhe, al2020multi, morshed2020cpu, dai2014accelerating, wang2013fpga, agrawal2023fab, cousins2016designing, cousins2014fpga, syafalni2022efficient, jayet2015polynomial,feldmann2021f1, samardzic2022craterlake, yang2023poseidon, ozturk2016custom, kim2022bts, nabeel2023cofhee, cao2014high} explore 
domain-specific architectures, such as GPUs, FPGAs, and ASICs, to accelerate homomorphic operations. These efforts have achieved significant speedups compared to CPUs. However, challenges remain in resource limitations, data movement, and practical implementation of \mkk{especially} ASIC-based accelerators \cite{kim2022bts}.}

In this work, our \textbf{goal} is to evaluate the suitability of real-world general-purpose processing-in-memory architectures to compute homomorphic operations. 
To this end, we implement homomorphic addition and multiplication on the UPMEM PIM \mkk{system} \cite{upmemsdk, gomez2022benchmarking, gomez2021benchmarking}, and evaluate them on real-world statistical and machine learning workloads. 


\juan{Processing-in-memory (PIM)~\cite{gomez2021benchmarking, gomez2022benchmarking, kim2021aquabolt, kim2022aquabolt, lee2022improving, ke2021near, upmemsdk, gomez2022machine, item2023transpimlibispass, kim2014hbm, ghose.bookchapter19, oliveira2021damov, devaux2019true, ghose2019processing, mutlu2022modern, seshadri2017ambit, stone1970logic, Kautz1969, farmahini2014drama, singh2018review, kwon2019tensordimm, yitbarek2016exploring, gupta2022memfhe, hajinazar2021simdram, lim2023design} \guptah{systems} can accelerate memory-intensive applications~\cite{item2023transpimlibispass, diab2023aim, giannoula2022sigmetrics, gomez2022experimental, oliveira2022accelerating} by equipping memory arrays with compute capabilities.} \guptah{These systems can potentially address the challenge of large ciphertexts in HE \hgg{algorithms} by reducing the overhead of data transfers between \guptahhh{the}
memory and the CPU \cite{mutlu2019processing, gomez2022machine}. In addition to reducing data movement, PIM also offers high levels of parallelism \cite{gomez2021benchmarking, gomez2022benchmarking}, which are useful for performing
costly \mkk{homomorphic} operations.} \mkk{Thus, by computing directly in memory, PIM can significantly improve the performance of HE.}
\mkk{Various} real-world PIM \mkk{systems have recently \hgg{been introduced}
}~\cite{gomez2021benchmarking, gomez2022benchmarking, kim2021aquabolt, kim2022aquabolt, lee2022improving, ke2021near, upmemsdk, gomez2022machine, item2023transpimlibispass, kim2014hbm, devaux2019true}. 
These real-world PIM systems have some common characteristics~\cite{gomez2022experimental}: \juan{there is a central host processor connected to conventional main memory, alongside PIM-enabled memory chips with processing elements that access memory with high bandwidth and low latency.} 
In this work, we \juan{use the} UPMEM PIM \mkk{system}~\cite{upmemsdk, gomez2022benchmarking, gomez2021benchmarking, guomachine}, 
\juan{which consist\guptah{s} of fine-grained multithreaded PIM cores near DRAM banks.} 
\juan{For more details \mkk{on the UPMEM PIM system}, we refer the reader to~\cite{ghose.bookchapter19, oliveira2021damov, devaux2019true, ghose2019processing, mutlu2022modern, seshadri2017ambit, stone1970logic, Kautz1969, farmahini2014drama, singh2018review, kwon2019tensordimm, yitbarek2016exploring, gupta2022memfhe, hajinazar2021simdram, upmemsdk, gomez2022benchmarking, gomez2021benchmarking}.}

\section{Implementation} \label{sec:implementation}

\juan{We consider an environment where users offload computations on encrypted data \guptah{to a PIM system}. Users handle key generation, encryption, and decryption to guarantee their data privacy. Computation of homomorphic operations takes place 
\guptah{in} a PIM system. In this work, we implement addition and multiplication operations.} 



\hg{The security level of HE relies on the polynomial modulus degree~\cite{sealcrypto}, affecting ciphertext length, vulnerability to attacks, and noise tolerance. For instance, for 27-bit security, we need a polynomial that has 1024 27-bit coefficients, \guptah{which} indicates a relatively lower security level in HE. Increasing the bit length enhances security.}
In this work, we also \guptahh{evaluate} 54-bit (2048-coefficient polynomial) and 109-bit (4096-coefficient polynomial) security levels. 
To represent 27-, 54-, and 109-bit coefficients, we use integers of 32, 64, and 128 bits, respectively. The reason is that the UPMEM PIM \mkk{system} that we use in our evaluation has native support for 32-bit integers.

\noindent
\juan{\textbf{\guptah{Homomorphic Addition. }}
We implement \guptah{homomorphic} 
addition \guptah{using polynomial addition}\cite{sontag1985real, zippel1993effective}  on the UPMEM PIM \mkk{system}. 
Each PIM thread running on a PIM core performs the element-wise addition of the coefficients of two polynomials. 
UPMEM PIM cores~\cite{upmemsdk} support native 32-bit integer addition (\texttt{add}) and 32-bit integer addition with carry-in (\texttt{addc}), which we use to implement 64- and 128-bit addition (and can be extended to any multiple of 32 bits).
}



\noindent
\juan{\textbf{\guptah{Homomorphic Multiplication.}} 
We implement \guptah{homomorphic} 
multiplication \guptah{using polynomial multiplication and polynomial addition}~\cite{moenck1976practical, harvey2017faster, harvey2022polynomial, chen2014high}. Each PIM thread running on a PIM core performs the \guptah{polynomial multiplication and polynomial addition of the coefficients of two polynomials to generate the desired result}. \guptah{For} 32-bit coefficients, we rely on the compiler-generated 32-bit shift-and-add based multiplication.\guptahh{\footnote{The UPMEM PIM \mkk{system} performs 8-bit and 16-bit multiplications using the native 8-bit \mkk{hardware} multipliers, but employs a \mkk{software-based} shift-and-add algorithm for higher bit widths~\cite{upmemsdk, gomez2022experimental, gomez2021benchmarking}.}}
For 64- and 128-bit multiplications, we \guptah{divide the bits into chunks of 32-bits and} apply the Karatsuba algorithm~\cite{eyupoglu2015performance}, which requires less operations than the traditional multiplication algorithm. 
We do not incorporate Number Theoretic Transform (NTT)\cite{bisheh2021high, fritzmann2019efficient} techniques to optimize multiplication. We leave them for future work.
}


\noindent
\guptah{\textbf{Statistical Workloads.}
We implement three statistical workloads (arithmetic mean, variance, linear regression) using homomorphic addition and homomorphic multiplication techniques. The arithmetic mean \cite{jacquier2003geometric, zhao2019approximating} \mkk{workload} employs polynomial addition performed \mkk{on the UPMEM PIM cores} and
scalar division \hggg{performed on} the \guptahhh{host processor.} 
\mkk{The} variance \cite{larson2008analysis, davidian1987variance} \mkk{workload uses} polynomial multiplication \mkk{which is performed on the UPMEM PIM cores} and 
\hgg{a} final scalar division \hggg{performed on}
the \mkk{host processor}. \mkk{Similarly, }linear regression \cite{su2012linear, maulud2020review} \mkk{workload uses} both polynomial addition and multiplication \mkk{to perform the vector-matrix multiplication}, \mkk{which} is employed \mkk{on the UPMEM PIM cores}.}

\section{Evaluation} \label{sec:evaluation}

\subsection{Methodology}

\vspace{-1mm}

\juan{We evaluate homomorphic addition and multiplication on a \guptah{first-generation} UPMEM PIM system \cite{upmemsdk, gomez2022benchmarking, gomez2021benchmarking, guomachine}, a 4-core Intel i5-8250U CPU \cite{i5-8250U}, and an NVIDIA A100 GPU\cite{a100}. 
The UPMEM system contains 2,524 PIM cores (running at 425 MHz) and 158GB of \guptahhh{PIM-enabled} 
memory with a total bandwidth of 2,145 GB/s. 
We compare our PIM implementations to \guptah{our own} custom CPU and GPU implementations. We also compare to an optimized CPU implementation, the SEAL CPU library~\cite{sealcrypto}, which leverages \guptah{the} Residue Number System (RNS)~\cite{gomathisankaran2011horns} and \guptahhh{the} Number Theoretic Transform (NTT)~\cite{mohsen2018performance} \guptahh{implementations} for faster operations.}


\juan{We first evaluate microbenchmarks \guptah{for} vector addition and vector multiplication (Section~\ref{sec:microbenchmarks}). We experiment with different numbers of ciphertexts between 20,480 to 327,680 for addition, and between 5,120 and 81,920 for multiplication. We run experiments for integers of 32 bits (27-bit coefficients), 64 bits (54-bit coefficients), and 128 bits (109-bit coefficients). 
We then evaluate SHE implementations of three statistical workloads (arithmetic mean, variance, linear regression) that employ our \guptahhh{PIM-based} homomorphic \guptah{encryption} operations (Section~\ref{sec:workloads}).} \mkk{We plan to open-source all workloads.}

\subsection{\juan{Vector Addition and Multiplication}}

\vspace{-1mm}

\begin{figure}[h]
    \centering
    \subfigure[\textbf{128-bit ciphertext vector addition}
    ]{
        \includegraphics[width=0.4\textwidth]{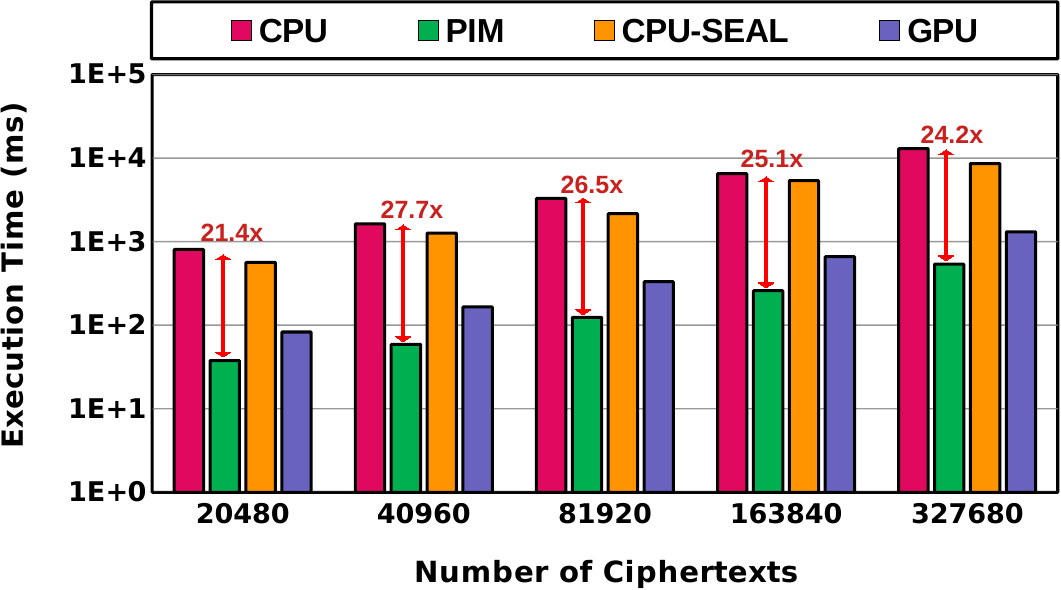}
        \label{fig:subfig2-add128}
    }
    \subfigure[\textbf{128-bit ciphertext vector multiplication}
    ]{
        \includegraphics[width=0.4\textwidth]
        {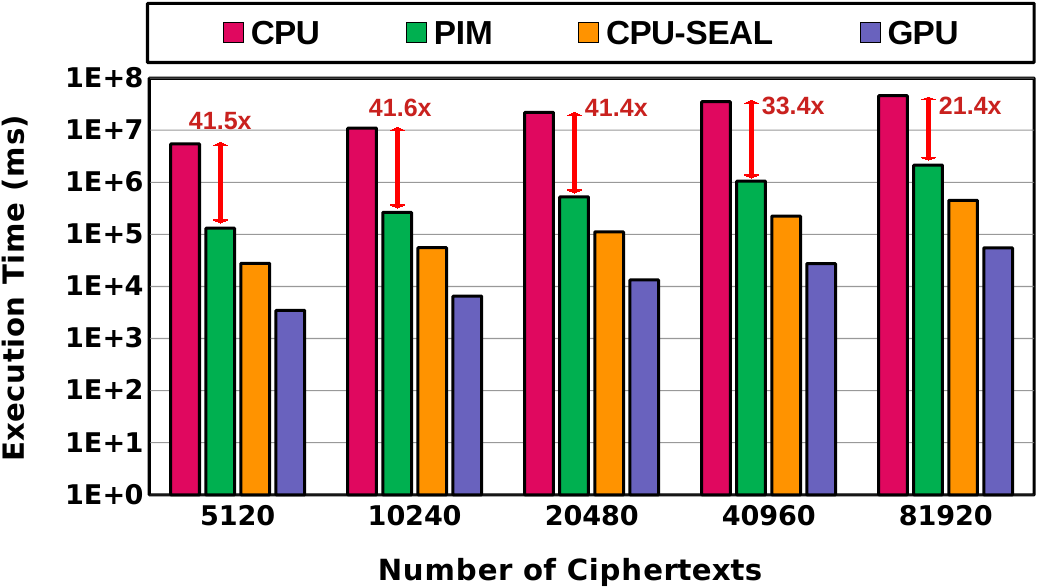}
        \label{fig:subfig2-mul128}
    }

    \vspace{-2mm}
    \caption{
    \juan{Execution time (ms) of ciphertext vector addition (a) and \guptah{vector} multiplication (b)
    \mkk{for 128-bit (109-bit) wide polynomial coefficients
    on CPU, PIM, CPU-SEAL and GPU.}} }
    \label{fig:main_figure_add1}
    \vspace{-2mm}
\end{figure}

\label{sec:microbenchmarks}
\juan{
Figure~\ref{fig:main_figure_add1} shows the execution time of vector addition (\ref{fig:subfig2-add128}) and multiplication (\ref{fig:subfig2-mul128}) \guptah{on homomorphically encrypted ciphertexts} for our \guptah{real-world UPMEM PIM-based} implementation (PIM), \guptahh{our} custom CPU and GPU implementations, and the SEAL library (CPU-SEAL). 
The figure also shows the speedup of PIM over the custom CPU \guptah{implementations}.} 

\juan{We make several observations about these experimental results. First, the performance of PIM implementations saturates at 11 or more PIM threads (not shown in Figure~\ref{fig:main_figure_add}). 
This is in line with the observations in prior works~\cite{gomez2021benchmarking, gomez2022machine, gomez2022experimental}.} \juan{Second, the large number of PIM cores and the native support \guptah{for} 32-bit integer addition \guptah{in PIM cores} result in fast execution of vector addition on the PIM system. 
Figure~\ref{fig:subfig2-add128} shows the results for 128-bit addition. The trends are the same for 32-bit and 64-bit addition. 
For 32-, 64-, and 128-bit addition, the PIM implementation outperforms CPU, CPU-SEAL, and GPU by $20-150\times$, $35-80\times$, and $15-50\times$, respectively.} 

\takeaway{\juan{With native \mkk{hardware} support for 32-bit integer addition and large number of PIM cores, the UPMEM PIM system outperforms CPU and GPU for homomorphic addition.}}

\juan{Third, vector multiplication on the UPMEM PIM system suffers from the lack of native 32-bit multiplication hardware, as multiplication \guptah{wider than \mkk{16 bits}} is based on \guptah{compiler generated} shift-and-add algorithm.  
Figure~\ref{fig:subfig2-mul128} shows the results for 128-bit multiplication. We observe similar trends for 32-bit and 64-bit multiplication. 
For 32-, 64-, and 128-bit multiplication, the PIM implementation outperforms CPU by $40-50\times$, and CPU-SEAL for 32 bits by $2\times$. However, \guptah{the PIM implementation} is $12-15\times$ slower than GPU, and $2-4\times$ slower than CPU-SEAL for 64 and 128 bits.}

\takeaway{\juan{The lack of native support for 32-bit integer multiplication hampers the performance of PIM for homomorphic multiplication. Future PIM systems with native 32-bit multiplication hardware \guptah{could} potentially outperform CPUs and GPUs.}}




\vspace{-2mm}

\subsection{\juan{Statistical Workloads}}
\label{sec:workloads}



\begin{figure}[h]
    \centering
    \subfigure[\textbf{Arithmetic Mean} 
    ]{
        \includegraphics[width=0.40\textwidth]{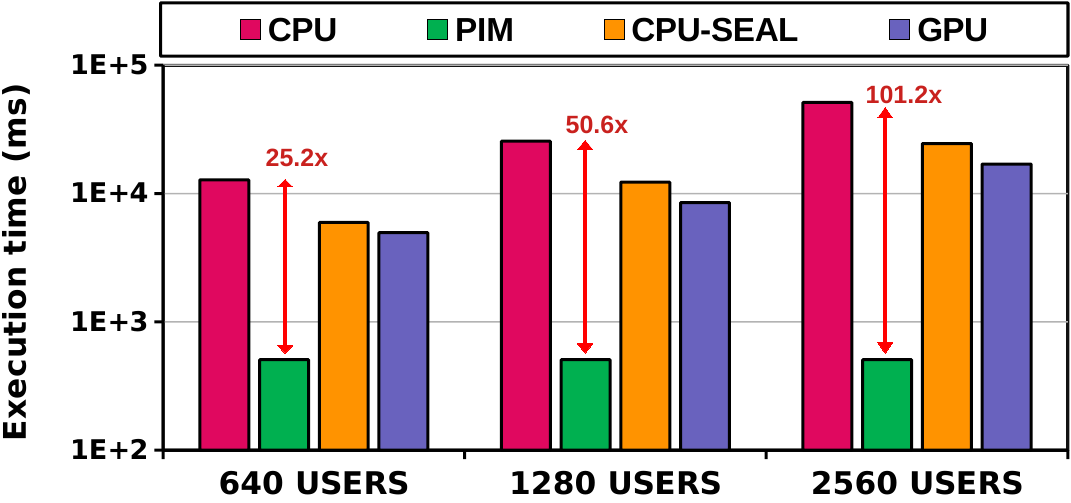}
        \label{fig:mean_subfig2}
    }\\
    \subfigure[\textbf{Variance} 
    ]{
        \includegraphics[width=0.40\textwidth]{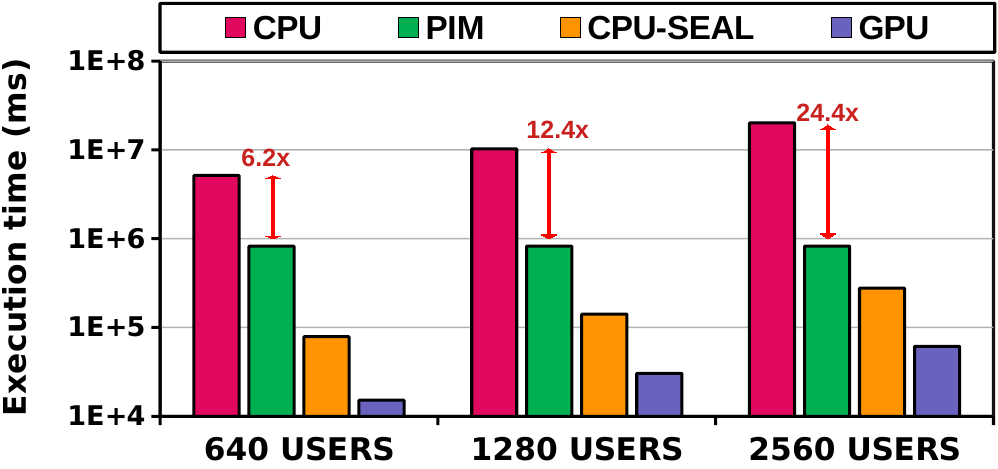}
        \label{fig:var_subfig2}
    }\\
    \subfigure[\textbf{Linear Regression} 
    ]{
        \includegraphics[width=0.42\textwidth]{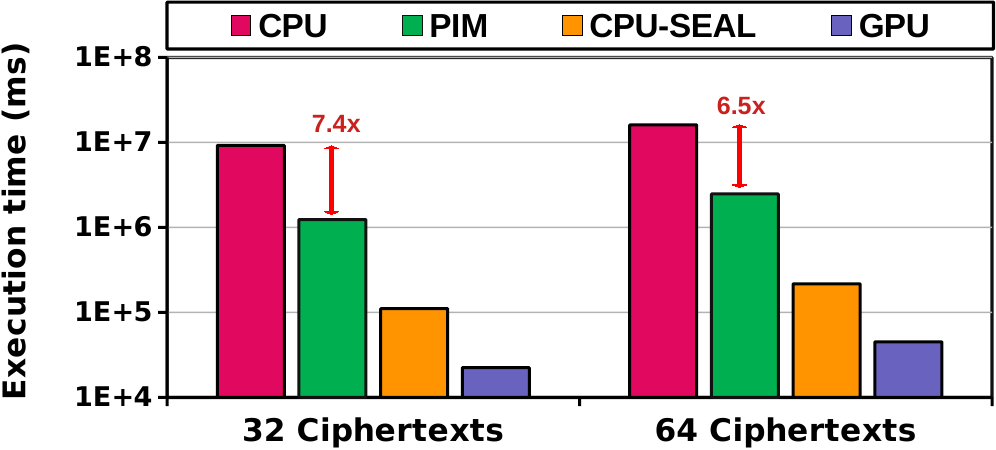}
        \label{fig:lr_subfig2}
    }\\
    
\vspace{-2mm}
    \caption{
    \juan{Execution time (ms) of aritmetic mean (a), variance (b), and linear regression (c)
    \mkk{for 128-bit (109-bit) wide polynomial coefficients
    on CPU, PIM, CPU-SEAL and GPU. }}}
    \label{fig:main_figure_add}
\vspace{-2mm}
\end{figure}

\juan{We \guptah{implement and} evaluate the performance of three real-world statistical workloads (arithmetic mean, variance, linear regression) that utilize homomorphic addition and multiplication \guptah{for the CPU, real-world PIM, CPU-SEAL \mkk{and GPU implementations}}. 
Figure~\ref{fig:main_figure_add} shows \guptahh{the} execution times of the three workloads on CPU, \mkk{PIM}, CPU-SEAL, and GPU. 
For arithmetic mean and variance, we evaluate scenarios with 640, 1280, and 2560 users. 
For linear regression, we \guptahh{evaluate} 640 users, and 32 and 64 ciphertexts per user (data samples with 3 features).
}

We make several observations from Figure~\ref{fig:main_figure_add}.
\juan{First, arithmetic mean uses only homomorphic addition. As a result, PIM is significantly faster than CPU, CPU-SEAL, and GPU. Figure~\ref{fig:mean_subfig2} shows PIM speedups of $25-100\times$ over CPU, $11-50\times$ over CPU-SEAL, and $9-34\times$ over GPU for different \guptah{number\mkk{s} of} users.} \juan{Second, as variance uses \guptah{the} square operation (i.e., \guptah{homomorphic} multiplication of two equal numbers), the PIM implementation is heavily burdened by the slow multiplication. In Figure~\ref{fig:var_subfig2}, we observe that PIM outperforms \guptah{only} the custom CPU implementation (by $6-25\times$) for different \guptahhh{numbers of} users. CPU-SEAL and GPU are, respectively, $2-10\times$ and $13-50\times$ faster than PIM.} \juan{Third, for linear regression the trends are the same as for variance, given that \guptah{linear regression also} uses multiplication heavily. Figure~\ref{fig:lr_subfig2} shows that PIM is only faster than the custom CPU implementation (by $7.5\times$) for 32 ciphertexts. CPU-SEAL and GPU are, respectively, $11.4\times$ and $54.9\times$ faster than PIM for 64 ciphertexts.} 
\mkk{Fourth, we observe that PIM execution time remains constant for different numbers of users. This is achieved by dynamically adjusting the utilization of PIM cores, which is particularly beneficial in our experiments as they involve a large number of users. This approach differs from CPUs and GPUs, which have a limited number of cores and must \hgg{use} 
them regardless of the number of users in our experiment.}

\vspace{0mm}



\takeaway{\guptah{The computational power of PIM scales with memory capacity~\cite{ahn2015scalable, ahn2023retrospective} \mkk{via}
the addition of more memory banks and corresponding PIM cores. \hgg{This memory-capacity-proportional performance}
scalability \hgg{provided by PIM} holds promise for accommodating expanding numbers of users 
\guptahhh{and} 
more parallel computations \hggg{as memory capacity grows}.}}






\vspace{-0.3em}
\section{Related Work} \label{sec:other_related_works}

\juan{Several recent works explore the suitability of real-world processing-in-memory (PIM) architectures~\cite{gomez2021benchmarking, gomez2022benchmarking, kim2021aquabolt, kim2022aquabolt, lee2022improving, ke2021near, upmemsdk, gomez2022machine, item2023transpimlibispass, kim2014hbm, ghose.bookchapter19, oliveira2021damov, devaux2019true, ghose2019processing, mutlu2022modern, seshadri2017ambit, stone1970logic, Kautz1969, farmahini2014drama, singh2018review, kwon2019tensordimm, yitbarek2016exploring, gupta2022memfhe, hajinazar2021simdram, lim2023design} to accelerate a variety of memory-intensive tasks~\cite{item2023transpimlibispass, diab2023aim, giannoula2022sigmetrics, gomez2022experimental, oliveira2022accelerating}. 
\guptah{T}o our knowledge, \guptah{this} 
is the first work to explore the 
\guptah{use} of \hgg{a} real PIM \mkk{system} \guptah{to accelerate}
homomorphic \guptahh{operations}.}

\juan{Acceleration of homomorphic operations on GPU\guptah{s}, FPGA\guptah{s}, or ASIC\guptah{s} is the subject of \guptah{various} recent works. \guptah{All these processor-centric techniques \mkk{suffer from} data movement bottlenecks} \mkk{between memory and compute units.}
GPUs can accelerate HE schemes~\cite{al2020towards, dai2016cuhe, al2020multi, morshed2020cpu, dai2014accelerating}. However, GPUs suffer from high power consumption for \hgg{homomorphic} operations~\cite{tan2021cryptgpu, wang2013exploring}. 
FPGAs can also accelerate \hgg{homomorphic }operations~\cite{wang2013fpga, agrawal2023fab, cousins2016designing, cousins2014fpga, syafalni2022efficient, jayet2015polynomial}, but they are limited in hardware resources and suffer from data movement \guptah{bottlenecks}~\cite{roy2018hepcloud, roy2019fpga}. 
}\guptahh{Several recent works propose ASIC designs ~\cite{feldmann2021f1, samardzic2022craterlake, yang2023poseidon, ozturk2016custom, kim2022bts, nabeel2023cofhee, cao2014high} for CKKS algorithms, but they are only evaluated in simulation. Similarly, PIM-based solutions~\cite{gupta2021accelerating, gupta2022memfhe, nejatollahi2020cryptopim} for accelerating homomorphic operations are also limited to simulation.}
\section{Conclusion}


\juang{\mk{We presented initial results on the use} of \hggg{a} real-world general-purpose PIM 
\hgg{architecture} 
(
\hgg{i.e., the} UPMEM PIM \mkk{system}~\cite{devaux2019true, gomez2021benchmarking}) to accelerate \mkk{homomorphic operations.} 
Our PIM implementations of homomorphic \hgg{addition, multiplication} and statistical workloads (mean, variance, linear regression) show great promise when compared to CPU and GPU implementations, as long as the necessary integer operations are natively supported by the PIM hardware. 
We \mkk{aim to} implement more homomorphic operations and optimizations as future work.}

\section*{Acknowledgments}
We acknowledge support from the SAFARI Research Group’s industrial partners, especially Google, Huawei, Intel, Microsoft, VMware, and the Semiconductor Research Corporation. 
This research was partially supported by the ETH Future Computing Laboratory and the European Union's Horizon programme for research and innovation under grant agreement No. 101047160, project BioPIM (Processing-in-memory architectures and programming libraries for bioinformatics algorithms). 
This research was also partially supported by ACCESS – AI Chip Center for Emerging Smart Systems, sponsored by InnoHK funding, Hong Kong SAR.

\bibliographystyle{IEEEtran}

\bibliography{refs}

\begin{thebibliography}{100}
\providecommand{\url}[1]{#1}
\csname url@samestyle\endcsname
\providecommand{\newblock}{\relax}
\providecommand{\bibinfo}[2]{#2}
\providecommand{\BIBentrySTDinterwordspacing}{\spaceskip=0pt\relax}
\providecommand{\BIBentryALTinterwordstretchfactor}{4}
\providecommand{\BIBentryALTinterwordspacing}{\spaceskip=\fontdimen2\font plus
\BIBentryALTinterwordstretchfactor\fontdimen3\font minus \fontdimen4\font\relax}
\providecommand{\BIBforeignlanguage}[2]{{%
\expandafter\ifx\csname l@#1\endcsname\relax
\typeout{** WARNING: IEEEtran.bst: No hyphenation pattern has been}%
\typeout{** loaded for the language `#1'. Using the pattern for}%
\typeout{** the default language instead.}%
\else
\language=\csname l@#1\endcsname
\fi
#2}}
\providecommand{\BIBdecl}{\relax}
\BIBdecl

\bibitem{ogburn2013homomorphic}
M.~Ogburn \emph{et~al.}, ``{Homomorphic Encryption},'' \emph{Procedia Computer Science}, 2013.

\bibitem{tourky2016homomorphic}
D.~Tourky \emph{et~al.}, ``{Homomorphic Encryption The “Holy Grail” of Cryptography},'' in \emph{ICCC 2016}.

\bibitem{gentry2011implementing}
C.~Gentry and S.~Halevi, ``{Implementing Gentry’s Fully Homomorphic Encryption Scheme},'' in \emph{EUROCRYPT 2011}.

\bibitem{al2020towards}
A.~Al~Badawi \emph{et~al.}, ``{Towards The Alexnet Moment For Homomorphic Encryption: HCNN, The First Homomorphic CNN on Encrypted Data With GPUs},'' \emph{TETC 2020}.

\bibitem{gentry2009fully}
C.~Gentry, ``{Fully Homomorphic Encryption using Ideal Lattices},'' in \emph{STOC 2009}.

\bibitem{van2010fully}
M.~Van~Dijk \emph{et~al.}, ``{Fully Homomorphic Encryption over the Integers},'' in \emph{EUROCRYPT 2010}.

\bibitem{cryptoeprint:2014/356}
D.~Boneh \emph{et~al.}, ``{Fully Key-Homomorphic Encryption, Arithmetic Circuit ABE, and Compact Garbled Circuits},'' IACR 2014.

\bibitem{boneh2013private}
D.~Boneh \emph{et~al.}, ``{Private Database Queries using Somewhat Homomorphic Encryption},'' in \emph{ACNS}, 2013.

\bibitem{moore2014practical}
C.~Moore \emph{et~al.}, ``{Practical Homomorphic Encryption: A Survey},'' in \emph{ISCAS 2014}.

\bibitem{chaudhary2019Analysis}
P.~Chaudhary \emph{et~al.}, ``{Analysis and Comparison of Various Fully Homomorphic Encryption Techniques},'' in \emph{GUCON 2019}.

\bibitem{gupta2021accelerating}
S.~Gupta and T.~{\v{S}}. Rosing, ``{Accelerating Fully Homomorphic Encryption with Processing-in-memory},'' in \emph{DAC 2021}.

\bibitem{samardzic2021f1}
N.~Samardzic \emph{et~al.}, ``{F1: A Fast and Programmable Accelerator for Fully Homomorphic Encryption},'' in \emph{MICRO 2021}.

\bibitem{samardzic2022craterlake}
N.~Samardzic \emph{et~al.}, ``{Craterlake: A Hardware Accelerator for Efficient Unbounded Computation on Encrypted Data},'' in \emph{ISCA 2022}.

\bibitem{alaya2020homomorphic}
B.~Alaya \emph{et~al.}, ``{Homomorphic Encryption Systems Statement: Trends and Challenges},'' \emph{CSR 2020}.

\bibitem{el2015challenges}
K.~El~Makkaoui \emph{et~al.}, ``{Challenges of Using Homomorphic Encryption to Secure Cloud Computing},'' in \emph{CloudTech 2015}.

\bibitem{gupta2022memfhe}
S.~Gupta \emph{et~al.}, ``{MemFHE: End-to-end Computing with Fully Homomorphic Encryption in Memory},'' \emph{TECS 2022}.

\bibitem{cao2015optimised}
X.~Cao \emph{et~al.}, ``{Optimised Multiplication Architectures For Accelerating Fully Homomorphic Encryption},'' \emph{TC 2015}.

\bibitem{su2022highly}
Y.~Su \emph{et~al.}, ``{A Highly Unified Reconfigurable Multicore Architecture to Speed-up NTT/INTT for Homomorphic Polynomial Multiplication},'' \emph{TVLSI 2022}.

\bibitem{doroz2016homomorphic}
Y.~Dor{\"o}z \emph{et~al.}, ``{Homomorphic AES Evaluation using the Modified LTV Scheme},'' \emph{DCC 2016}.

\bibitem{mert2019design}
A.~C. Mert \emph{et~al.}, ``{Design and Implementation of Encryption/Decryption Architectures for BFV Homomorphic Encryption Scheme},'' \emph{TVLSI 2019}.

\bibitem{meftah2022towards}
S.~Meftah \emph{et~al.}, ``{Towards High Performance Homomorphic Encryption for Inference Tasks on CPU: An MPI Approach},'' \emph{FGCS 2022}.

\bibitem{morshed2020cpu}
T.~Morshed \emph{et~al.}, ``{CPU and GPU Accelerated Fully Homomorphic Encryption},'' in \emph{HOST 2020}.

\bibitem{dai2016cuhe}
W.~Dai and B.~Sunar, ``{cuHE: A Homomorphic Encryption Accelerator Library},'' in \emph{BalkanCryptSec 2016}.

\bibitem{al2020multi}
A.~Al~Badawi \emph{et~al.}, ``{Multi-GPU Design and Performance Evaluation of Homomorphic Encryption on GPU Clusters},'' \emph{TPDS 2020}.

\bibitem{dai2014accelerating}
W.~Dai \emph{et~al.}, ``{Accelerating NTRU based Homomorphic Encryption using GPUs},'' in \emph{HPEC 2014}.

\bibitem{wang2013fpga}
W.~Wang and H.~Xinming, ``{FPGA Implementation Of a Large-number Multiplier for Fully Homomorphic Encryption},'' in \emph{ISCAS 2013}.

\bibitem{agrawal2023fab}
R.~Agrawal \emph{et~al.}, ``{FAB: An FPGA-based Accelerator for Bootstrappable Fully Homomorphic Encryption},'' in \emph{HPCA 2023}.

\bibitem{cousins2016designing}
D.~B. Cousins \emph{et~al.}, ``{Designing an FPGA-accelerated Homomorphic Encryption Co-processor},'' \emph{TETC 2016}.

\bibitem{cousins2014fpga}
D.~B. Cousins \emph{et~al.}, ``{An FPGA Co-processor Implementation of Homomorphic Encryption},'' in \emph{HPEC 2014}.

\bibitem{syafalni2022efficient}
I.~Syafalni \emph{et~al.}, ``{Efficient Homomorphic Encryption Accelerator with Integrated PRNG using Low-cost FPGA},'' \emph{IEEE Access 2022}.

\bibitem{jayet2015polynomial}
C.~Jayet-Griffon \emph{et~al.}, ``{Polynomial Multipliers for Fully Homomorphic Encryption on FPGA},'' in \emph{ReConFig 2015}.

\bibitem{feldmann2021f1}
A.~Feldmann \emph{et~al.}, ``{F1: A Fast and Programmable Accelerator for Fully Homomorphic Encryption},'' \emph{MICRO 2021}.

\bibitem{yang2023poseidon}
Y.~Yang \emph{et~al.}, ``{Poseidon: Practical Homomorphic Encryption Accelerator},'' in \emph{HPCA 2023}.

\bibitem{ozturk2016custom}
E.~{\"O}zt{\"u}rk \emph{et~al.}, ``{A Custom Accelerator for Homomorphic Encryption Applications},'' \emph{TC 2016}.

\bibitem{kim2022bts}
S.~Kim \emph{et~al.}, ``{BTS: An Accelerator for Bootstrappable Fully Homomorphic Encryption},'' in \emph{ISCA 2022}.

\bibitem{nabeel2023cofhee}
M.~Nabeel \emph{et~al.}, ``{CoFHEE: A Co-processor for Fully Homomorphic Encryption Execution},'' in \emph{DATE 2023}.

\bibitem{cao2014high}
X.~Cao \emph{et~al.}, ``{High-speed Fully Homomorphic Encryption over the Integers},'' in \emph{FC 2014}.

\bibitem{gomez2021benchmarking}
J.~Gómez-Luna \emph{et~al.}, ``{Benchmarking Memory-centric Computing Systems: Analysis of Real Processing-in-Memory Hardware},'' in \emph{IGSC 2021}.

\bibitem{gomez2022benchmarking}
J.~G{\'o}mez-Luna \emph{et~al.}, ``{Benchmarking a New Paradigm: Experimental Analysis and Characterization of a Real Processing-in-memory System},'' \emph{IEEE Access 2022}.

\bibitem{kim2021aquabolt}
J.~H. Kim \emph{et~al.}, ``{Aquabolt-XL: Samsung HBM2-PIM with In-memory Processing for ML Accelerators and Beyond},'' in \emph{HCS 2021}.

\bibitem{kim2022aquabolt}
J.~H. Kim \emph{et~al.}, ``{Aquabolt-XL HBM2-PIM, LPDDR5-PIM with In-memory Processing, and AXDIMM with Acceleration Buffer},'' \emph{IEEE MICRO 2022}.

\bibitem{lee2022improving}
D.~Lee \emph{et~al.}, ``{Improving In-Memory Database Operations with Acceleration DIMM (AxDIMM)},'' in \emph{DaMoN}, 2022.

\bibitem{ke2021near}
L.~Ke \emph{et~al.}, ``{Near-Memory Processing in Action: Accelerating Personalized Recommendation with AxDIMM},'' \emph{IEEE Micro}, 2021.

\bibitem{upmemsdk}
``{UPMEM SDK},'' \url{https://sdk.upmem.com/2023.1.0/}.

\bibitem{gomez2022machine}
J.~G{\'o}mez-Luna \emph{et~al.}, ``{Machine Learning Training on a Real Processing-in-Memory System},'' in \emph{ISVLSI 2022}.

\bibitem{item2023transpimlibispass}
M.~Item \emph{et~al.}, ``{TransPimLib: Efficient Transcendental Functions for Processing-in-Memory Systems},'' in \emph{ISPASS}, 2023.

\bibitem{kim2014hbm}
J.~Kim and Y.~Kim, ``{HBM: Memory Solution for Bandwidth-hungry Processors},'' in \emph{HCS}, 2014.

\bibitem{ghose.bookchapter19}
S.~Ghose \emph{et~al.}, ``{The Processing-in-Memory Paradigm: Mechanisms to Enable Adoption},'' in \emph{Beyond-CMOS Technologies for Next Generation Computer Design 2019}.

\bibitem{oliveira2021damov}
G.~F. Oliveira \emph{et~al.}, ``{DAMOV: A New Methodology And Benchmark Suite For Evaluating Data Movement Bottlenecks},'' \emph{IEEE Access 2021}.

\bibitem{devaux2019true}
F.~Devaux, ``{The True Processing in Memory Accelerator},'' in \emph{HCS}, 2019.

\bibitem{ghose2019processing}
S.~Ghose \emph{et~al.}, ``{Processing-in-memory: A Workload-driven Perspective},'' \emph{IBM JRD 2019}.

\bibitem{mutlu2022modern}
O.~Mutlu \emph{et~al.}, ``{A Modern Primer on Processing-in-memory},'' in \emph{Emerging Computing: From Devices to Systems: Looking Beyond Moore and Von Neumann 2022}.

\bibitem{seshadri2017ambit}
V.~Seshadri \emph{et~al.}, ``{Ambit: In-memory Accelerator for Bulk Bitwise Operations Using Commodity DRAM Technology},'' in \emph{MICRO 2017}.

\bibitem{stone1970logic}
H.~S. Stone, ``{A Logic-in-Memory Computer},'' \emph{IEEE TC}, 1970.

\bibitem{Kautz1969}
W.~H. {Kautz}, ``{Cellular Logic-in-Memory Arrays},'' \emph{IEEE TC}, 1969.

\bibitem{farmahini2014drama}
A.~Farmahini-Farahani \emph{et~al.}, ``{DRAMA: An Architecture for Accelerated Processing-near-memory},'' \emph{IEEE Computer Architecture Letters}, 2014.

\bibitem{singh2018review}
G.~Singh \emph{et~al.}, ``{A Review of Near-memory Computing Architectures: Opportunities and Challenges},'' in \emph{Euromicro DSD 2018}.

\bibitem{kwon2019tensordimm}
Y.~Kwon \emph{et~al.}, ``{TensorDIMM: A Practical Near-memory Processing Architecture for Embeddings and Tensor Operations in Deep Learning},'' in \emph{MICRO 2019}.

\bibitem{yitbarek2016exploring}
S.~F. Yitbarek \emph{et~al.}, ``{Exploring Specialized Near-memory Processing for Data-intensive Operations},'' in \emph{DATE 2016}.

\bibitem{hajinazar2021simdram}
N.~Hajinazar \emph{et~al.}, ``{SIMDRAM: A Framework for Bit-serial SIMD Processing-using-DRAM},'' in \emph{ASPLOS 2021}.

\bibitem{lim2023design}
C.~Lim \emph{et~al.}, ``{Design and Analysis of a Processing-in-DIMM Join Algorithm: A Case Study with UPMEM DIMMs},'' \emph{ACM SIGMOD}, 2023.

\bibitem{li2019leveraging}
W.~Li \emph{et~al.}, ``{Leveraging Memory PUFs and PIM-based Encryption to Secure Edge Deep Learning Systems},'' in \emph{VTS 2019}.

\bibitem{reis2020computing}
D.~Reis \emph{et~al.}, ``{Computing-in-memory for Performance and Energy-efficient Homomorphic Encryption},'' \emph{TVLSI 2020}.

\bibitem{gomez2022experimental}
J.~G{\'o}mez-Luna \emph{et~al.}, ``{An Experimental Evaluation of Machine Learning Training on a Real Processing-in-Memory System},'' \emph{arXiv preprint arXiv:2207.07886}, 2022.

\bibitem{halevi2019improved}
S.~Halevi \emph{et~al.}, ``{An Improved RNS Variant of the BFV Homomorphic Encryption Scheme},'' in \emph{CT-RSA 2019}.

\bibitem{wibawa2022bfv}
F.~Wibawa \emph{et~al.}, ``{BFV-Based Homomorphic Encryption for Privacy-Preserving CNN Models},'' \emph{Cryptography 2022}.

\bibitem{mono2022finding}
J.~Mono \emph{et~al.}, ``{Finding and Evaluating Parameters for BGV},'' \emph{Cryptology ePrint Archive}, 2022.

\bibitem{cheon2020remark}
J.~H. Cheon \emph{et~al.}, ``{Remark on the Security of CKKS Scheme in Practice},'' \emph{Cryptology ePrint Archive}, 2020.

\bibitem{acar2018survey}
A.~Acar \emph{et~al.}, ``{A Survey on Homomorphic Encryption Schemes: Theory and Implementation},'' \emph{ACM CSUR 2018}.

\bibitem{alexandru2020cloud}
A.~B. Alexandru \emph{et~al.}, ``{Cloud-based Quadratic Optimization with Partially Homomorphic Encryption},'' \emph{IEEE TAC}, 2020.

\bibitem{damgaard2012multiparty}
I.~Damg{\aa}rd \emph{et~al.}, ``{Multiparty Computation From Somewhat Homomorphic Encryption},'' in \emph{CRYPTO 2012}.

\bibitem{yasuda2014practical}
M.~Yasuda \emph{et~al.}, ``{Practical Packing Method in Somewhat Homomorphic Encryption},'' in \emph{DPM 2013 and SETOP 2013}.

\bibitem{de2021does}
L.~de~Castro \emph{et~al.}, ``{Does Fully Homomorphic Encryption Need Compute Acceleration?}'' \emph{IACR 2021}.

\bibitem{diab2023aim}
S.~Diab \emph{et~al.}, ``{A Framework for High-throughput Sequence Alignment using Real Processing-in-Memory Systems},'' \emph{Bioinformatics}, 2023.

\bibitem{giannoula2022sigmetrics}
C.~Giannoula \emph{et~al.}, ``{Towards Efficient Sparse Matrix Vector Multiplication on Real Processing-in-Memory Architectures},'' in \emph{SIGMETRICS}, 2022.

\bibitem{oliveira2022accelerating}
G.~F. Oliveira \emph{et~al.}, ``{Accelerating Neural Network Inference with Processing-in-DRAM: From the Edge to the Cloud},'' \emph{IEEE Micro 2022}.

\bibitem{mutlu2019processing}
O.~Mutlu \emph{et~al.}, ``{Processing Data Where It Makes Sense: Enabling In-memory Computation},'' \emph{MICPRO}, 2019.

\bibitem{guomachine}
J.~Gómez-Luna \emph{et~al.}, ``{Evaluating Machine Learning Workloads on Memory-Centric Computing Systems},'' in \emph{ISPASS 2023}.

\bibitem{sealcrypto}
``{Microsoft SEAL},'' \url{https://www.microsoft.com/en-us/research/project/microsoft-seal/}.

\bibitem{sontag1985real}
E.~D. Sontag, ``{Real Addition and the Polynomial Hierarchy},'' \emph{IPL 1985}.

\bibitem{zippel1993effective}
R.~Zippel, \emph{{"Effective Polynomial Computation"}}.\hskip 1em plus 0.5em minus 0.4em\relax SSBM 1993.

\bibitem{moenck1976practical}
R.~T. Moenck, ``{Practical Fast Polynomial Multiplication},'' in \emph{SYMSAC 1976}.

\bibitem{harvey2017faster}
D.~Harvey \emph{et~al.}, ``{Faster Polynomial Multiplication over Finite Fields},'' \emph{JACM 2017}.

\bibitem{harvey2022polynomial}
D.~Harvey \emph{et~al.}, ``{Polynomial multiplication over finite fields in time},'' \emph{JACM 2022}.

\bibitem{chen2014high}
D.~D. Chen \emph{et~al.}, ``{High-speed Polynomial Multiplication Architecture for Ring-LWE and SHE Cryptosystems},'' \emph{TCAS-I 2014}.

\bibitem{eyupoglu2015performance}
C.~Eyupoglu, ``{Performance Analysis of Karatsuba Multiplication Algorithm for Different Bit Lengths},'' \emph{Procedia: SBS 2015}.

\bibitem{bisheh2021high}
M.~Bisheh-Niasar \emph{et~al.}, ``{High-Speed NTT-based Polynomial Multiplication Accelerator For CRYSTALS-Kyber Post-Quantum Cryptography},'' \emph{ICAR 2021}.

\bibitem{fritzmann2019efficient}
T.~Fritzmann and J.~Sep{\'u}lveda, ``{Efficient and Flexible Low-Power NTT for Lattice-Based Cryptography},'' in \emph{HOST 2019}.

\bibitem{jacquier2003geometric}
E.~Jacquier \emph{et~al.}, ``{Geometric or Arithmetic Mean: A Reconsideration},'' \emph{FAJ}, 2003.

\bibitem{zhao2019approximating}
T.-H. Zhao \emph{et~al.}, ``{On Approximating the Quasi-arithmetic Mean},'' \emph{JIA}, 2019.

\bibitem{larson2008analysis}
M.~G. Larson, ``{Analysis of Variance},'' \emph{Circulation}, 2008.

\bibitem{davidian1987variance}
M.~Davidian \emph{et~al.}, ``{Variance Function Estimation},'' \emph{JASA}, 1987.

\bibitem{su2012linear}
X.~Su \emph{et~al.}, ``{Linear Regression},'' \emph{WIREs Comp Stats}, 2012.

\bibitem{maulud2020review}
D.~Maulud \emph{et~al.}, ``{A Review on Linear Regression Comprehensive in Machine Learning},'' \emph{JASTT}, 2020.

\bibitem{i5-8250U}
Intel, ``{Intel® Core™ i5-8250U Processor},'' \url{https://ark.intel.com/content/www/us/en/ark/products/124967/intel-core-i58250u-processor-6m-cache-up-to-3-40-ghz.html}, 2017.

\bibitem{a100}
{NVIDIA}, ``{NVIDIA A100 Tensor Core GPU Architecture. White Paper},'' \url{https://images.nvidia.com/aem-dam/en-zz/Solutions/data-center/nvidia-ampere-architecture-whitepaper.pdf}, 2020.

\bibitem{gomathisankaran2011horns}
M.~Gomathisankaran \emph{et~al.}, ``{HORNS: A Homomorphic Encryption Scheme for Cloud Computing using Residue Number System},'' in \emph{IEEE CISS 2011}.

\bibitem{mohsen2018performance}
A.~W. Mohsen \emph{et~al.}, ``{Performance Analysis of Number Theoretic Transform for Lattice-based Cryptography},'' in \emph{ICCES 2018}.

\bibitem{ahn2015scalable}
J.~Ahn \emph{et~al.}, ``{A Scalable Processing-in-memory Accelerator for Parallel Graph Processing},'' in \emph{ISCA}, 2015.

\bibitem{ahn2023retrospective}
J.~Ahn \emph{et~al.}, ``{Retrospective: A Scalable Processing-in-memory Accelerator for Parallel Graph Processing},'' \emph{arXiv preprint arXiv:2306.15577}, 2023.

\bibitem{tan2021cryptgpu}
S.~Tan \emph{et~al.}, ``{CryptGPU: Fast Privacy-preserving Machine Learning on the GPU},'' in \emph{SP 2021}.

\bibitem{wang2013exploring}
W.~Wang \emph{et~al.}, ``{Exploring the Feasibility of Fully Homomorphic Encryption},'' \emph{IEEE TC 2013}.

\bibitem{roy2018hepcloud}
S.~S. Roy \emph{et~al.}, ``{HEPCloud: An FPGA-based Multicore Processor for FV Somewhat Homomorphic Function Evaluation},'' \emph{TC 2018}.

\bibitem{roy2019fpga}
S.~S. Roy \emph{et~al.}, ``{FPGA-based High-performance Parallel Architecture For Homomorphic Computing on Encrypted Data},'' in \emph{HPCA 2019}.

\bibitem{nejatollahi2020cryptopim}
H.~Nejatollahi \emph{et~al.}, ``{CryptoPIM: In-memory Acceleration for Lattice-based Cryptographic Hardware},'' in \emph{DAC 2020}.

\end{thebibliography}





\end{document}